\documentclass[preprint,showpacs,preprintnumbers,amsmath,amssymb]{revtex4}
\usepackage{graphicx}% Include figure files
\usepackage{dcolumn}% Align table columns on decimal point
\usepackage{bm}% bold math

\newcommand{\ds}{\displaystyle}

\newcommand{\be}{\begin{equation}}
\newcommand{\en}{\end{equation}}
\newcommand{\bea}{\begin{eqnarray}}
\newcommand{\ena}{\end{eqnarray}}
\begin{document}

\title{Closed inflationary universe in Patch Cosmology}
%\author{ Sergio del Campo, Ram\'on
%Herrera, Pedro Labra{\~n}a and Joel Saavedra}
%\address{Instituto de F\'{\i}sica, Pontificia Universidad Cat\'olica de
%Valpara\'{\i}so, Casilla 4950, Valpara\'{\i}so, Chile.}
\author{Sergio del Campo\footnote{Electronic Mail-address:
sdelcamp@ucv.cl}} \affiliation{Instituto de F\'{\i}sica,
Pontificia Universidad Cat\'{o}lica de Valpara\'{\i}so, Avenida
Brasil 2950, Casilla 4059, Valpara\'{\i}so, Chile.}
\author{Ram\'{o}n Herrera\footnote{E-mail address: ramon.herrera@ucv.cl}}
\affiliation{Instituto de F\'{\i}sica, Pontificia Universidad
Cat\'{o}lica de Valpara\'{\i}so, Avenida Brasil 2950, Casilla
4059, Valpara\'{\i}so, Chile.}
\author{Pedro Labra{\~n}a\footnote{E-mail address: plabrana@ubiobio.cl}}
\affiliation{Departamento de F\'{\i}sica, Universidad del
B\'{\i}o–B\'{\i}o, Avenida Collao 1202, Casilla 5-C, Concepci\'on,
Chile.}
\author{Joel Saavedra\footnote{E-mail address: joel.saavedra@ucv.cl}}
\affiliation{Instituto de F\'{\i}sica, Pontificia Universidad
Cat\'{o}lica de Valpara\'{\i}so, Avenida Brasil 2950, Casilla
4059, Valpara\'{\i}so, Chile.}
\date{\today}

\begin{abstract}

 In this article we study  closed
inflationary universe models using the Gauss-Bonnet Brane. We
determine and characterize the existence of a universe with
$\Omega > 1$, with an appropriate period of inflation. We have
found that this model is less restrictive in comparison with the
 standard approach where a scalar field is considered. We use recent
astronomical observations  to constrain  the parameters appearing
in the model.

\end{abstract}

\maketitle

\section{\label{sec:level1} Introduction}

Cosmological inflation has become an integral part of the standard
model of the universe. Apart from being capable of removing the
shortcomings of the standard cosmology, it gives important clues
for structure formation in the universe. The scheme of inflation
\cite{IC} (see \cite{libro} for a review) is based on the idea
that at early times there was a phase in which the universe
evolved through accelerated expansion in a short period of time at
high energy scales. During this phase, the universe was dominated
by a potential $V(\phi)$ of a scalar field $\phi$, which is called
the inflaton.

Normally, inflation has been associated with a flat universe, due
to its ability to effectively drive the spatial curvature  to
zero. In fact, requiring sufficient inflation to homogenize random
initial conditions, drives the universe very close to its critical
density. In this context, the recent observations are entirely
consistent with a universe having a total energy density  very
close to its critical value and having an almost scale invariant
power spectrum \cite{Peiris:2003ff, WMAP3}. Most people interpret
these values as corresponding to a flat universe. But, according
to this results, we might take the alternative point of view of a
marginally open \cite{op1,op} or closed \cite{Linde:2003hc, Joel,
Cl} universe, which at early times in its evolution presents an
inflationary period of expansion.

%%%%%%%%%%%%%%%%%%%%%%%%%%%%%%%%%%%%%%%%%%%%%%%%%%%%%%%%%%%%%%%%%%%%%%

Nowadays there is considerable interest in inflationary models
motivated by string/M-theory (for a review  see Refs.\cite{1}). In
particular, much attention has been focused on the brane world
scenario (BW), where our observable four-dimensional universe is
modelled as a domain wall embedded in a higher-dimensional bulk
space \cite{2}.

BW cosmology offers a novel approach to our understanding of the
evolution of the universe, the most spectacular consequence of
this scenario is the modification of the Friedmann equation. These
kind of models can be obtained from superstring
theory~\cite{witten, witten2}. For a comprehensible review on BW
cosmology, see Refs.~\cite{lecturer,lecturer2,lecturer3}.
Specifical consequences of a chaotic inflationary universe
scenario in a BW model were previously described~\cite{maartens},
where it was found that the slow-roll approximation is enhanced by
the modification of the Friedmann equation.

On the other hand, when a stage closed to the Big-Bang is studied,
quantum effects should be included in the bulk. In the high
dimensional theory, high curvature correction terms should be
added to the Einstein Hilbert action. One of the correction term
to this action  is the so called Gauss-Bonnet (GB) combination.
The GB term arises naturally as the leading order of the $\alpha'$
expansion of heterotic superstring theory, where, $\alpha'$ is the
inverse string tension. In fact, all versions of string theory
(except Type II) in $10$ dimensions include this term. Therefore,
it could be interesting to make study the effects of the
Gauss-Bonnet term on inflationary braneworld models
\cite{Meng:2003pn,Lidsey:2003sj, Calcagni:2004as, Kim:2004gs,
Kim:2004jc, Kim:2004hu, Murray:2006fw}.

 When the five dimensional Einstein-GB equations are projected on
to the brane, a complicated Hubble equation  is obtained \cite{Ch,
Davis:2002gn, Gravanis:2002wy}. Interestingly enough, this
modified Friedmann equation reduces to a very simple equation
$H^2\sim \rho^q$ with $q = 1, 2, 2/3$ corresponding to General
Relativity (GR), Randall Sundrum (RS) and GB regimes respectively.
This situation motivated the "patch cosmology" as a useful
approach to study braneworld scenarios \cite{Calcagni:2004bh}.
This scheme makes use of a nonstandard Friedmann equation of the
form $H^2= \kappa_q \rho^q$. Despite all the shortcomings of this
approximate treatment of extra-dimensional physics, it gives
several important first-impact information.

The purpose of the present work is to study closed inflationary
universe models in the spirit of Linde's work \cite{Linde:2003hc},
where the matter content is confined to a four dimensional brane
which is embedded in a five dimensional bulk where a Gauss-Bonnet
(GB) contribution is considered. We study these models using the
approach of patch cosmology.

%%%%%%%%%%%%%%%%%%%%%%%%%%%%%

The paper is organized as follows. In Sect.~\ref{Sec1} we briefly
review the cosmological equations in the Gauss-Bonnet brane world
and present the patch cosmological equations for this model. In
Sect. III we determine the characteristics of a closed
inflationary universe model with a constant scalar potential.
Also, we get the value of the scalar field, when inflation begins
and we obtain the probability of the creation of a close universe
from nothing. In Sect. IV we consider a chaotic inflationary
model. In Sect. V the cosmological perturbations are investigated.
Finally, in Sect. VI, we summarize our results.

\section{\label{Sec1} Cosmological Equations in Gauss-Bonnet brane}

We start with the five-dimensional bulk action for the
Gauss-Bonnet braneworld:

\begin{eqnarray}
S &=&\frac{1}{2\kappa
_{5}^{2}}\int_{bulk}d^{5}x\sqrt{-g_{5}}\left\{ R-2\Lambda
_{5}+\alpha \left( R^{\mu \nu \lambda \rho }R_{^{\mu \nu \lambda
\rho }}-4R^{\mu \nu }R_{\nu \mu }+R^{2}\right) \right\} \nonumber \\
&&+\int_{brane}d^{4}x\sqrt{-g_{4}}\left(
\mathcal{L}_{matter}-\sigma \right),\label{action1}
\end{eqnarray}
where $\Lambda _{5}=-3\mu ^{2}\left( 2-4\alpha \mu ^{2}\right) $
is the cosmological constant in five dimensions, with the
$AdS_{5}$ energy scale $\mu$, $\alpha$ is the GB coupling
constant, $\kappa _{5}=8\pi/m_{5}$ is the five dimensional
gravitational coupling constant and $\sigma$ is the brane tension.
For a Friedmann-Robertson-Walker (FRW) metric, the exact
Friedmann-like equation becomes \cite{Ch, Davis:2002gn,
Gravanis:2002wy}

\begin{equation}
2\mu \sqrt{1+\frac{H^{2}}{\mu ^{2}}}\left( 3-4\alpha \mu
^{2}+8\alpha H^{2}\right)+\frac{k}{a^{2}} =\kappa _{5}^{2}\left(
\rho +\sigma \right) ,  \label{q3}
\end{equation}
where $\rho$ represents the energy density of the matter sources
on the brane, $a$ is the scale factor, and $k=0,+1,-1$ represents
a flat, closed or open spatial section, respectively. The modified
Friedmann equation (\ref{q3}) encodes all cosmological
information. Despite the rather complicated form of Eq.
(\ref{q3}), it is possible to make progress if  we use the
dimensionless variable $\chi$ \cite{Lidsey:2003sj},
\begin{equation}
\label{var1} \kappa_5^2(\rho+\sigma) =
\left[{{2(1-4\alpha\mu^2)^3} \over {\alpha} }\right]^{1/2}
\sinh\chi\,,
\end{equation}
The Friedmann equation can be written as
 \begin{equation}
H^2 = {1\over
4\alpha}\left[(1-4\alpha\mu^2)\cosh\left({2\chi\over3}
\right)-1\right]\,,\label{q22}
 \end{equation}
where $\chi$ represents a dimensionless measure of the energy
density. The modified Friedman equation~(\ref{q22}), together with
Eq.~(\ref{var1}), ensures the existence of one characteristic
Gauss-Bonnet energy scale,
 \be \label{gbscale}
m_\alpha= \left[{{2(1-4\alpha\mu^2)^3} \over {\alpha} \kappa_5^4
}\right]^{1/8}\,,
 \en
such that the GB high energy regime ($\chi\gg1$) occurs if
$\rho+\sigma \gg m_\alpha^4$.  Expanding Eq.~(\ref{q22}) in $\chi$
and using (\ref{var1}), we find in the full theory three regimes
for the dynamical history of the brane universe,

$\bullet$\,\,Gauss-Bonnet regime (5D),
 \be
\rho\gg m_\alpha^4~ \Rightarrow ~ H^2\approx \left[ {\kappa_5^2
\over 16\alpha}\, \rho \right]^{2/3}\,,\label{gbl}
 \en
$\bullet$\,\, Randall-Sundrum regime (5D),
 \be
 m_\alpha^4 \gg
\rho\gg\sigma \equiv m_\sigma^4 ~ \Rightarrow ~ H^2\approx
{\kappa_4^2 \over 6\sigma}\, \rho^{2}\,,\label{rsl}
 \en
$\bullet$\,\, Einstein-Hilbert regime (4D),
 \be
\rho\ll\sigma~ \Rightarrow ~ H^2\approx {\kappa_4^2 \over 3}\,
\rho\,. \label{ehl}
 \en
Clearly Eqs. (\ref{gbl}), (\ref{rsl}) and (\ref{ehl}) are much
simpler than the full Eq (\ref{q3}) and in a practical case one of
the three energy regimes will be assumed. Therefore, patch
cosmology can be useful to describe the universe in a region of
time and energy in which \cite{Calcagni:2004bh}
\begin{equation}
H^{2}=\kappa_{q}\rho ^{q}-\frac{1}{a^{2}}, \label{dda}
\end{equation}
where $H=\dot{a}/a$ is the Hubble parameter  and $q$ is a patch
parameter that describes a particular cosmological model under
consideration. The choice $q=1$ corresponds to the standard
General Relativity with $\kappa_1=8\pi/3m_{p}^2$, where $m_{p}$ is
the four dimensional Planck mass. If we take $q=2$, we obtain the
high energy limit of the brane world cosmology, in which
$\kappa_2=4\pi/3\sigma m_p^2 $. Finally, for $q=2/3$, we have the
GB brane world cosmology, with $\kappa_{2/3}=G_{5}/16\zeta$, where
$G_5$ is the $5D$ gravitational coupling constant and
$\zeta=1/8g_s$ is the GB coupling ($g_s$ is the string energy
scale). The parameter $q$, which describes the effective degrees
of freedom from gravity, can take a value in a non-standard set
because of the introduction of non-perturbative stringy
effects.Just to mention some possibilities, these are the presence
of a complicated geometrical framework with either compact and
non-compact extra dimensions, multiple and/or folding branes
configurations, and so on. For instance, in Ref.\cite{Kim:2004hu}
it was found that an appropriate region to a patch parameter $q$
is given by $1/2 = q < \infty$. On the other hand, from Cardassian
cosmology it is possible to obtain a Friedmann equation like
(\ref{dda}) as a consequence of embedding our observable universe
as a 3+1 dimensional brane in extra dimensions. In fact, in
Ref.\cite{Chung:1999zs} a modified FRW equation  was obtained in
our observable brane with $H^2 \sim \rho^n$ for any $n$. This
result was obtained using five-dimensional Einstein equations plus
the Israel boundary conditions that related the energy-momentum on
the brane to the derivatives of the metric in the bulk.

 Brane world models are characterized by the feature that
standard model matter is confined to a 1+3 dimensional brane while
gravity propagates in the higher dimensional bulk. In general
terms, we should include the situation in which matter could be
transferred from the bulk to the brane or vice-versa. This
circumstance could be realizable only if the bulk contains an
appropriate form of matter, expressed by a high dimensional
component of the energy-momentum tensor. However, in this paper we
will assume that the matter fields are restricted to a lower
dimensional hypersurface (brane) and that gravity exists
throughout the space-time (brane and bulk) as a dynamical theory
of geometry. Also, for 4D homogeneous and isotropic (FRW)
cosmology, an extended version of Birkhoff's theorem tells us that
if the bulk space-time is AdS, it
 implies that  the effect of the Weyl tensor (known as dark radiation)
does not appear in the modified Friedmann equation \cite{Bi}.
Certainly,  it could be interesting to consider this effect, but
its study will be postponed for the present. Thus, the energy
conservation equation on the brane follows directly from the
Gauss-Codazzi equations. For a perfect fluid matter source it is
reduced to the familiar form,

\begin{equation}
\dot{\rho}+3H\left( \rho +P\right) =0,  \label{q4}
\end{equation}
where $\rho$ and $P$ represent the energy and pressure densities,
respectively. The dot denotes derivative with respect to the
cosmological time $t$.

 We consider that the matter content of the universe is a homogeneous
inflaton field $\phi(t)$ with potential $V(\phi)$. Then the energy
density and pressure are given by

\begin{equation}
\rho=\frac{\dot{\phi}^2}{2}+V(\phi) = P + 2 V(\phi),\label{phii}
\end{equation}
and the energy conservation equation (\ref{q4}) becomes
\begin{equation}
\displaystyle \ddot{\phi}+3\,H\,\dot{\phi}+V_{,\phi }(\phi
)=0\,\,. \label{ecphi}
\end{equation}

From the effective Friedmann equation (\ref{dda}) we can obtain
the equation of motion for the scale factor, \be
\ddot{a}=a\kappa_q
\rho^{q-1}\left[\rho\left(1-\frac{3}{2}q\right)-\frac{3}{2}\,q\,P\right],
\label{addot}
\end{equation}
and  for convenience we will use  units in which $c=\hbar=1$.

\section{Constant Potential in Gauss Bonnet Brane}
\label{sec1}

Following the scheme of Ref.~\cite{Linde:2003hc}, we study a
closed inflationary universe, where inflation is driven by a
single scalar field confined to the brane. First let us consider a
simple model with the step-like effective potential described by:
$V(\phi) = 0$ at $\phi< 0$; $V (\phi) = V_0 = Constant$ at $0<
\phi< \phi_0$ and $V(\phi)$ extremely steep for $\phi >\phi_0$. %%
We consider that the birth of the inflating closed universe can be
created "from nothing", in a state where the scalar field takes
the value $\phi_{in} \leq \phi_0$ at the point with $\dot{a}=0$,
$\dot{\phi}=0$  and the potential energy density in this point is
$V(\phi_{in})\gtrsim V_0$ . If the effective potential for $\phi <
\phi_0$ grows very sharply, then the scalar  field instantly falls
down to the value $\phi_0$, with potential energy $V(\phi_0)=
V_0$, and its initial potential energy
$V(\phi_{in})\equiv\,V_{in}$ becomes converted to the kinetic
energy. Since this process happens instantly we can consider
$\dot{a}=0$, so  that the scalar field arrives  to the plateau
with a velocity given by

 \be \ds
\dot{\phi}_0\,=\, -\sqrt{2\,(V_{in}-V_0)} \,\,\label{ec2b}, \en
where $\dot{\phi}_0$ is the initial velocity of the field $\phi$,
immediately after rolling down to the flat part of the potential.

Thus, in order to study inflation in this scenario, we need to
solve Eqs.~(\ref{dda}) and (\ref{ecphi}) in the interval $0<\phi
\leq \phi_0$, with initial conditions $\dot{\phi}=\dot{\phi}_0$,
$a=a_0$ and $\dot{a}=0$. These equations have different solutions,
depending on the value of $\dot{\phi}_0$ and the patch under
consideration. From Eqs. (\ref{dda}) and (\ref{ec2b}), we obtain

\be \frac{\ddot{a}}{a}
=\kappa_q\,V_{in}^{q-1}\left[V_{in}\left(1-3\,q\right)+3\,q\,V_0)
\right]\,\,\label{ec4}. \en

Then, we  notice that there are three different scenarios,
depending on the particular value of $V_{in}$. First, in the
particular case when \be \ds \frac{V_0}{V_{in}}\,=\,
\frac{3\;q-1}{3\;q} \label{sta}, \en we see that the initial
acceleration of the scale factor is $\ddot{a}=0$. Since initially
$\dot{a}=0$, then the universe remain static and the scalar field
moves with constant speed given by Eq.~(\ref{ec2b}).

In the second case we have

\be \ds \frac{V_0}{V_{in}}\,<\, \frac{3\;q-1}{3\;q}\label{col}.\en
In this case the universe start  moving with negative acceleration
($\ddot{a}<0$) from the state $\dot{a}=0$. Therefore, in the
scalar field equation the term proportional to $\dot{\phi}$
describe a negative dissipative term which makes the motion of the
field $\phi$ even faster, so that $\ddot{a}$ becomes more
negative. This universe rapidly collapses.

The third case corresponds to

\be \ds \frac{V_0}{V_{in}}\,>\, \frac{3\;q-1}{3\;q}\label{co2}.\en
Here we have $\ddot{a}>0$ and   the universe enters into an
inflationary stage.

Now we proceed to make a simple analysis of the cosmological Eqs.
(\ref{dda}) and (\ref{ecphi}) for cases where the condition
(\ref{co2}) is satisfied. In the interval $0<\phi < \phi_0$ the
inflaton field equation becomes

\be \ds \ddot{\phi}+ 3\,\frac{\dot{a}}{a}\,\dot{\phi} =
0\,\,\label{Vconst}, \en whose first integral is \be \ds
\dot{\phi}^2(t) =\dot{\phi_0}^2
\,\left[\frac{a_0}{a(t)}\right]^6\,\,\label{phidot}. \en

The behavior of the scalar field implies that the evolution of the
universe rapidly falls into an exponential regimen (inflationary
stage) where the scale factor becomes $a\sim\,e^{H\,t}$, and the
Hubble parameter for the patch cosmology reads as follows \be
H=\sqrt{\kappa_q\, V_0^q}. \en

Once the universe enters in the inflationary stage, the scalar
field moves by an amount  $\Delta \phi_{inf}$ and then stops. From
Eq.~(\ref{phidot}) we get

 \be \ds
\Delta \phi_{inf} = \frac{\dot{\phi_0}}{3 H}
\approx-\frac{1}{3}\sqrt{\frac{1}{\kappa_q\,V_0^{q-1}}}\label{fi-inf}.
\en Notice that, when $q= 1$ ($\kappa_{q=1}=\kappa_1= 8\pi/3$), we
obtain $\Delta \phi_{inf}\approx -\sqrt{\frac{1}{24\pi}}$, which
coincides with the result obtained in Ref.~\cite{Linde:2003hc},
and in the case $q=2$ ($\kappa_{q=2}=\kappa_2= 4\pi/3\;\sigma$),
we get $\Delta \phi_{inf}\approx
-\sqrt{\frac{\sigma}{12\pi\;V_{0}}}$, result which coincides  with
the high energy limit of  the model studied in Ref.\cite{Joel}.

In order to study the model at early times (i.e. before inflation
takes place), we write the equation of motion of the scale factor
Eq.~(\ref{dda}) in the following convenient form
 \be
\ddot{a}(t) = 2\kappa_1\, V_0\, a(t)\, \beta(t)\,\,\label{beta1},
\en
where, we have introduced a
 time-dependent dimensionless  parameter $\beta (t)$, defined as
 follows
 \be \beta(t) =
 \frac{1}{2}\,\frac{\kappa_q\;\rho^{q-1}}{\kappa_1\,V_0}\,\left[\rho
 (1-\frac{3}{2}\,q)-\frac{3}{2}\,q\,p\right] \label{def-bet}. \en

%We are going to consider that $\beta(0)\equiv\beta_0 \ll 1$

Now we proceed to make a  simple  analysis of the evolution of the
scalar field and scale factor before inflation. With this purpose
we consider initial conditions which satisfy
\mbox{$\beta(0)\equiv\beta_0 \ll 1$}. We can write

\be \beta_0 =
V_{in}^{q-1}\,[V_{in}-V_*]\,\frac{(1-3q)\,\kappa_q}{2\,V_0\,\kappa_1},
\en where we have defined $V_*=3qV_0/(3q-1)$.

Then, at the beginning of the process, we have $a(t) \approx a_0$
and $\beta(t) \approx \beta_0$, and Eq.(\ref{beta1}) takes the
form: \be \ddot{a}(t) = 2\kappa_1\, a_0 V_0
\beta_0\,\,\label{beta}. \en

For small $t$, the solution of this equation is
 \be \ds
a(t) = a_0 \left(1 + \kappa_1 \beta_0  V_0 t ^2\right).
\label{adet} \en

From Eqs. (\ref{phidot}) and (\ref{adet}) we find that the time
interval $\Delta t_1$  where $\beta(t)$ becomes twice as large as
$\beta_0$ is given by

\be \Delta
t_1\approx\sqrt{\frac{\kappa_1}{\kappa_q}}\;\left[4\pi\left(\frac{3}{2}\right)^{q-1}
\,[(3q-1)+(q-1)^2]\;V_0^{q}\right]^{-1/2} .\label{tiempo} \en

This result is in agreement with Refs. \cite{Linde:2003hc,Joel}
for the cases $q=1$ and $q=2$, ($\kappa_1/\kappa_2=2\sigma$)
respectively.

Consequently, the scalar field decreases  by the amount
 \be \ds
\Delta \phi_1 \sim  \dot{\phi_0} \, \Delta t_1 \approx
-\sqrt{\frac{\kappa_1}{\kappa_q}}\;\left[2\pi\left(\frac{3}{2}\right)^{q-1}
\,(3q-1)\;[(3q-1)+(q-1)^2]\;V_0^{q-1}\right]^{-1/2} \label{bexp}.
\en

We  note that for a given $q$, this result depends on the value of
$V_0$, and the decrease  of the scalar field  is less restrictive
than the one used in the standard case $q=1$ in which $\Delta
\phi_1 =const.\sim -1/(2\sqrt{3\pi})$ \cite{Linde:2003hc}.
After the time $\Delta t_2 \approx \Delta t_1$, the scalar field
decreases  by the amount $\Delta\phi_2 \approx \Delta\phi_1$, and
the rate of growth of $a(t)$ increases again. This process
finishes when $\beta(t) \rightarrow \beta_f$, where
$\beta_f=\kappa_q\,V_0^{q-1}/(2\kappa_1)$.

Since at each interval $\Delta t_i$ the value of $\beta$ doubles,
the number of intervals $n_{int}$ after which $\beta(t)
\rightarrow \beta_f$ is
 \be \ds
n_{int} =\frac{\ln \beta_f-\ln \beta_0}{\ln 2} \,\,\label{n}. \en

Therefore, if we know the initial value of the scalar field, we
can estimate the value of the  scalar field at which inflation
begins $\phi_{inf}$

\be \ds \label{ficero} \phi_{inf} \approx  \phi_0  - \left(
\frac{\ln \beta_f-\ln \beta_0}{\ln 2}
 \right)\!\sqrt{\frac{\kappa_1}{\kappa_q}}\;\left[2\pi\left(\frac{3}{2}\right)^{q-1}
\,(3q-1)\;[(3q-1)+(q-1)^2]\;V_0^{q-1}\right]^{-1/2}. \en  This
expression indicates that our result for $\phi_{inf}$ is sensitive
to the choice of the potential energy $V_0$, the patch parameter
$q$, and the initial velocity of the scalar field $\phi$
immediately after it rolls down to the plateau of the potential
energy.

Note that if the scalar field starts its motion with a small
velocity, inflation begins immediately. On the other hand, if the
field $\phi$ moves with a large initial velocity, inflation is
delayed.
After inflation, the field $\phi$ stops moving when it passes the
distance $|\Delta\phi_{\inf}|$ (see Eq. (\ref{fi-inf})). Note that
if the field stops before it reaches $\phi=0$, the universe
expands forever in an inflationary stage. The same problem occurs
in  Einstein's General Relativity model \cite{Linde:2003hc}.
However, in the context of patch cosmology ($q \neq 1$) the value
of $\Delta\phi_{\inf}$ depends on the value we assign to the patch
parameter and the other constants involved. Therefore, we will see
that the problem of the universe inflating forever disappears and
thus the inflaton field can reach the value $\phi =0$ for some
appropriate conditions of the constants involved in our models.

Since inflation could occur only in the interval $\phi_{inf} > 0$
and $\phi = 0$, the initial value of the inflaton field satisfies

\be \ds \label{condi}  \phi_0 >  \left( \frac{\ln \beta_f-\ln
\beta_0}{\ln 2}
 \right)\!\sqrt{\frac{\kappa_1}{\kappa_q}}\;\left[2\pi\left(\frac{3}{2}\right)^{q-1}
\,(3q-1)\;[(3q-1)+(q-1)^2]\;V_0^{q-1}\right]^{-1/2}. \en

On the other hand, in order to determine the initial value of the
scalar field $\phi_0$, we need to find the value of $\beta_0$. To
perform this task, we study the birth of a closed universe in the
patch cosmology. From the semiclassical point of view, the
probability of creation of a closed universe from nothing is given
by \cite{Koya}

\be P\sim e^{-2\mid S_E
\mid}=\exp\left(-\left[\frac{8\,\pi^2}{3\,\kappa_q^2}\right]\,\rho^{-(2q-1)}\right)\simeq\,
\exp\left(-\left[\frac{8\,\pi^2}{3\,\kappa_q^2}\right]\,V^{-(2q-1)}\right),
\en where we have used Eq.(\ref{phii}) with $\dot{\phi}^2\ll
V(\phi)$.

 We estimate the conditional probability in order  to create the universe
with an energy density equal to $V_{in}$. We  assume that this
energy  is smaller than $V_*$ so that it satisfies the condition
expressed by Eq.(\ref{co2}). We get

 \be \ds
P \sim e^{-2\mid S_E \mid} =
\exp\left(-\left[\frac{8\,\pi^2}{3\,\kappa_q^2}\right]\,\left[\frac{1}{V_{in}^{2q-1}}-\frac{1}{V_{*}^{2q-1}}\right]
\right)\,\,\label{prob}. \en

This expression tells us that the probability of creation of the
universe with $\beta_0 \neq 0$ is not exponentially suppressed if
$\beta_0 < \frac{6V_{0}}{m_{p}^{4}}$, $\beta_0 <
\frac{432}{625}\frac{V_{0}^{4}}{\sigma^{3}\,m_{p}^{4}}$, and
$\beta_0 <\frac{27\,m_p^2}{64\,\pi^2}\,\kappa_{2/3}^3$ for $q=1$,
$q=2$ and $q=2/3$, respectively. See table~(\ref{tab:table1}).
These conditions, together  with Eq.~(\ref{condi}), impose a bound
from below for the initial value of the scalar field $\phi_0$.

\begin{table}
\caption{\label{tab:table1}Probabilities of each patch cosmology
and the initial value of the inflaton field $\phi_0$.}
\begin{tabular}{||c||c||c||}
\hline\hline $q$ & Probability & $\,\phi _{0}/m_p >$ \\
\hline\hline
$1$ & $P\sim \exp \left( -\frac{\beta _{0}m_{p}^{4}}{6V_{0}}\right) $ & $6.57$ \\
\hline\hline $2$ & $P\sim \exp \left(- \frac{625}{432}\frac{\sigma
^{3}\beta _{0}m_{p}^{4}}{V_{0}^{4}}\right) $ & 11.10 \\
\hline\hline
$2/3$ & $P\sim \exp \left( -\frac{64\pi ^{2}}{27\,m_p^2}\frac{\beta _{0}}{%
k_{2/3}^{3}}\right) $ & 17.73 \\
\hline\hline
\end{tabular}
\end{table}

As an example, we  take a particular set of  the parameters
appearing  in our model. We consider $V_0 \sim 10^{-11} m_p^4$
\cite{libro}, $\kappa_1=8 \pi/3m_p^2$ , $\kappa_2=4
\pi/3\sigma\,m_p^2 $ where the brane tension was taken as $\sigma
=10^{-10}m_p^4$ and $\kappa_{2/3}=10^{-3}/m_p^{2/3}$. These values
allow us to fix the initial condition of the inflaton field for
the different patches. Table \ref{tab:table1} shows our results.

\begin{figure}[th]
\includegraphics[width=4.0in,angle=0,clip=true]{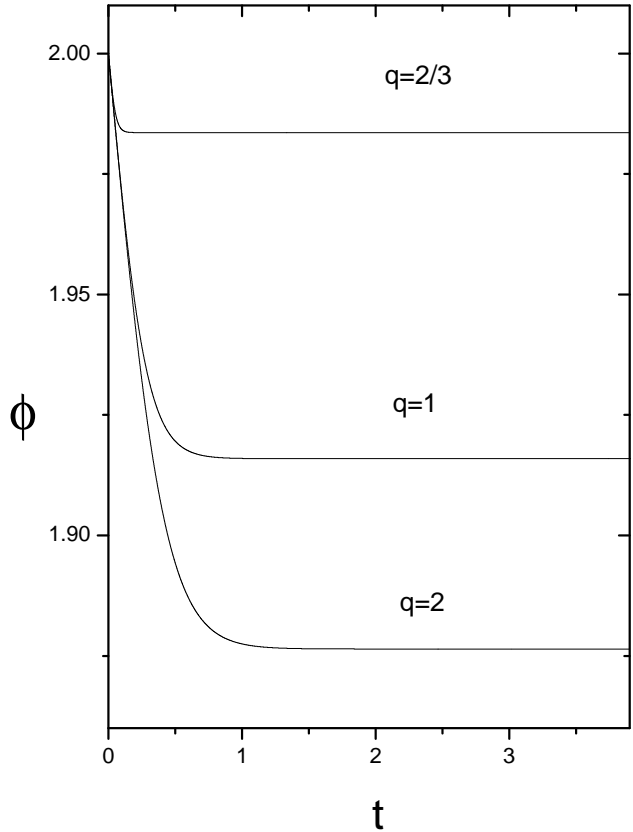}
\caption{ We plot the inflaton field $\phi$(t) as a function of
the cosmological time by using the model $V= V_0= const.$ in the
patch cosmology model. In all of them, we have taken the same
value of $\dot{\phi_0}$. Here, we have used unit where $m_p=1$.}
\label{fig1}
\end{figure}

Numerical solutions of the inflaton field $\phi(t)$ are shown in
Fig.~\ref{fig1}, where we have considered different values of the
patch parameter. Note that the interval from $\phi_0$ to
$\phi_{\inf}$ increases when the patch parameter $q$ increases,
but its shape remain practically unchanged.

We can notice that for these three values of $q$, inflation begins
immediately if the field $\phi$ starts its motion with
sufficiently small velocity, in analogy with Einstein's GR theory
($q=1$). On the other hand if it starts with large initial
velocity the universe could does not present the inflationary
period at any stage.

It is also worth mentioning, that in all patch cosmologies under
study the inflaton field does not show oscillations,since in this
case the scalar potential is a constant.

\section{Chaotic potential}
\label{sec2}

We consider an effective potential given by $\label{pot}
V(\phi)=\lambda_n\,\phi^n$ for $\phi < \phi_0$, which becomes
extremely steep for $\phi > \phi_0$. In particular, we take
$\lambda_n$ as a free parameter, which in the special case $n=2$
is related to the  mass of the scalar field
\cite{Fairbairn:2002yp}. In the following we will consider
$\lambda_n> 0$ and the cases $n= 2, 4$.
%%%%%%%%%%%%%%%%%%%%%%%%%%%%%%%%%%%%%%%%%%%%%%%%%%%%%%%%%%

We assume that the whole process is divided in three parts. The
first one corresponds to the creation of the closed universe
``from nothing" in a state where the inflaton field takes the
value $\phi_{in} \geq \phi_0$ at the point with $\dot{a}=0$,
$\dot{\phi}=0$, and where the potential energy is $V_{in}$. If the
effective potential for $\phi\ > \phi_0$ grows very sharply, then
the scalar field instantly falls down to the value $\phi_0$, with
potential energy $V(\phi_0)$. Therefore, the initial potential
energy becomes converted to kinetic energy, (see the previous
section). Then, we have

\be \dot{\phi}^2_0= 2(V_{in}-V(\phi_0)). \en

Following the discussion of the previous section we assume that
the following initial condition is satisfied:

\be \label{cond}  V_{in}< \frac{3q}{(3q-1)}\,V(\phi_0)\,, \en
which guarantees that the model arrives  to an inflationary
regimen. As it was mentioned previously, in all other cases the
universe remains either static, or it  collapses rapidly.

The following steps are described by Eqs.~(\ref{dda}) and
(\ref{ecphi})  in the interval $\phi \leq \phi_0$ with initial
conditions $\dot{\phi}=\dot{\phi}_0$, $a=a_0$ and $\dot{a}=0$. In
particular, the second part of the process corresponds to the
motion of the inflaton field before the beginning of the inflation
stage, and it is well described by the following approximated
field  equations.

\begin{equation}
\ddot{\phi} = -3\,\frac{\dot{a}}{a}\,\dot{\phi}\,, \label{t1}\\
\end{equation}
and
 \begin{equation}
 \label{t2} \ddot{a} =2
\kappa_q\,a\,V(\phi)\,\beta(t),
\end{equation}
with  $\beta(t)$ satisfying  $\beta(t)\ll 1$, as in the previous
section.

The last step corresponds to the stage of inflation where
$\dot{\phi}$ is small enough and the scale factor $a(t)$ grows
exponentially. This part is well described by the following
 equations

\be \ds 3\,\frac{\dot{a}}{a}\,\dot{\phi} =
-V_{,\,\phi}\,\,\label{inf1}, \en
\begin{equation}
\label{inf2} \ddot{a} = \kappa_q\,a\,V(\phi)^q\,.
\end{equation}

Now we will describe the process in  more details. Let us consider
the second stage, where the scalar field and the scale factor
$a(t)$ satisfy  Eqs.(\ref{t1}) and
 (\ref{t2}). Following the previous  scheme,
 we solve the equation for $a(t)$ by
considering $\beta(t) \ll 1$. Then, at the beginning of the
process, when $a\approx a_0$ and $\beta \approx \beta_0$,
Eq.~(\ref{t2}) takes the form

\be \ddot{a}(t) =2\kappa_1\, a_0 V(\phi_0)
\beta_0\,\,\label{beta}, \en and the scalar field satisfies
Eq.(\ref{phidot}). The amount of the decreasing of the scalar
field during the time $\Delta t$, which make the value of $\beta$
two times grater than $\beta_0$ is

\be \Delta \phi \approx
-\sqrt{\frac{\kappa_1}{\kappa_q}}\;\left[2\pi\left(\frac{3}{2}\right)^{q-1}
\,(3q-1)\;[(3q-1)+(q-1)^2]\;V(\phi_0)^{q-1}\right]^{-1/2}. \en

This process continues until $\dot{\phi}$ becomes small enough, so
that the universe begins to expand  in an exponential way,
characterizing the inflationary era. We take that the  inflation
begins when $\beta(t)$ tends to $\beta_f$, where
$\beta_f=\kappa_q\,V(\phi_0)^{q-1}/(2\kappa_1)$.  Then, according
to our previous result,  the  field $\phi$ gets the value:

\be \ds \label{inf3} \phi_{inf} \approx  \phi_0  - \left(
\frac{\ln \beta_f-\ln \beta_0}{\ln 2}
 \right)\!\sqrt{\frac{\kappa_1}{\kappa_q}}\;\left[2\pi\left(\frac{3}{2}\right)^{q-1}
\,(3q-1)\;[(3q-1)+(q-1)^2]\;V(\phi_0)^{q-1}\right]^{-1/2}. \en

During inflation the number of e-folds for our potential is given
by \be N \simeq
\frac{3\kappa_q\,\lambda_n^{q-1}}{n(n(q-1)+2)}\,\phi^{n(q-1)+2}_{inf}\,.
\en

We assume conclusively  that for $N=60$ one  would  have
$\Omega=1.1$. Then one can show that for $N=59.5$ and the same
value of the Hubble constant one would get $\Omega\approx 1.3$,
whereas for $N = 60.5$ we obtain $\Omega \approx 1.03$
\cite{Linde:2003hc,Cl}. Notice that, in analogy with Einstein's
theory of GR  a fine tuning of the value of $V(\phi_0)$  is
required for having the value of $\Omega$ in the range $1
\lesssim\Omega< 1.1$.

One of the main prediction of any inflationary universe model is
the primordial spectrum that arises due to quantum fluctuation of
the inflaton field. Therefore, it is interesting to study the
density perturbation behavior for our cosmology.

\section{\label{Sec3}  Perturbations }

Even though the study of scalar density perturbations in closed
universes is quite complicated, it is interesting to give an
estimation of the standard quantum scalar field fluctuations in
this scenario. In particular, the amplitude of scalar
perturbations generated during inflation for a flat space is
{\cite{Kim:2004gs}}
\be \delta_H^2 = \frac{1}{25\pi^2}\,\frac{H^2}{\dot{\phi}^2} =
\frac{1}{25\pi^2}\,\frac{9\kappa_q^3\,V^{3q}}{(V_{,\,\phi})^2}\,.
\label{ec10}
 \en

 \begin{figure}[th]
\includegraphics[width=4.0in,angle=0,clip=true]{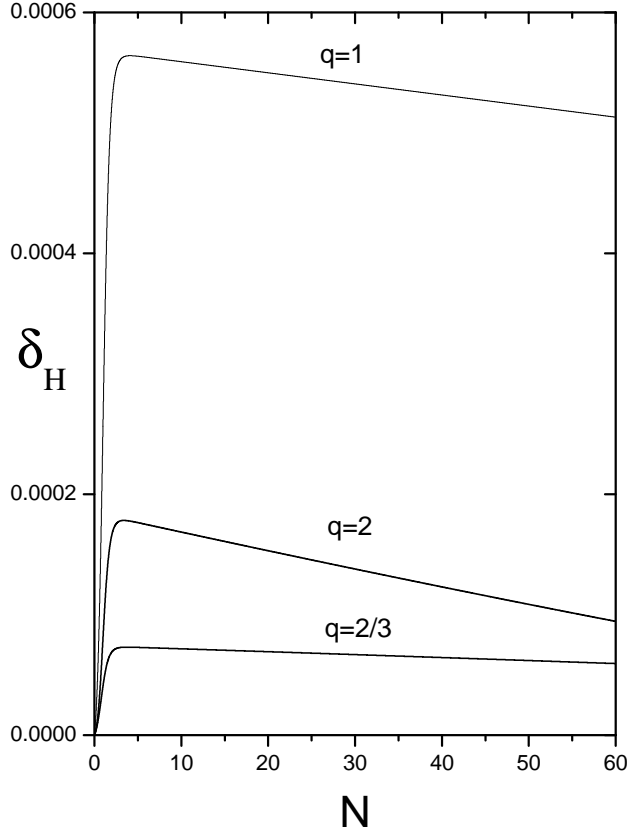}
 \caption{Scalar density perturbation  as a
function of the N e-folds. The plot shows the case $n=2$, for the
potential $V(\phi)$. We have taken $\sigma=10^{-10}$,
$\kappa_{2/3}=10^{-3}$ and we have used unit where $m_p=1$.}
\label{fig3}
\end{figure}

The slow-roll parameters are given by
\begin{eqnarray}
 \epsilon_1 =-\frac{\dot{H}}{H^2}&\simeq&\frac{q\,(V_{,\,\phi})^2}{6\kappa_q\,V^{1+q}}\,, \\
\epsilon_2 =-\frac{\ddot{H}}{H\,\dot{H}}&\simeq&
\frac{1}{3\kappa_q}\,\left[\frac{q}{2}\,\frac{(V_{,\,\phi})^2}{V^{1+q}}-\frac{V_{,\,\phi\phi}}{V^{q}}\right].\label{e2}
 \end{eqnarray}

Certainly, in our case, Eq.(\ref{ec10}) is an approximation and
must be supplemented by several different contributions in the
context of a closed inflationary universe \cite{Linde:2003hc}.
However, one may expect that the flat-space expression gives a
correct result for $N>3$.

If one interprets perturbations produced immediately after the
creation of a closed universe (at $N\sim O(1)$) as perturbations
on the horizon scale $l\sim 10^{28} cm$, then the maximum at
$N\sim 10$ would correspond to the scale $l\sim 10^{24} cm$, and
the maximum at $N\sim 15$ would correspond to the scale $l\sim
10^{22} cm$, which is similar to the galaxy scale.

We also consider the  spectral index $n_s$, which is related to
the power spectrum of density perturbations $\delta_H(k)$. For
modes with a wavelength much larger than the horizon ($k \ll a
H$), the spectral index $n_s$ is an exact power law, expressed by
$\delta_H(k) \propto k^{n_s-1}$, where $k$ is the comoving wave
number. The scalar spectral index is given by \cite{Kim:2004gs}
\begin{equation}
n_s=1-4\epsilon_1 - 2\epsilon_2\label{ns}.
\end{equation}

Following  Ref.{\cite{Kim:2004gs}} the running of the scalar
spectral index for our model becomes

\begin{equation}
\alpha_s=\frac{d n_s}{d\ln
k}\simeq\;-8\frac{\epsilon_1^2}{q}-10\epsilon_1\,\epsilon_2+2\epsilon_2^2-2\epsilon_3,\label{dnsdk}
\end{equation}
where we have used that $d\ln k=-dN$ and $\epsilon_3$ is given by
$$
\epsilon_3\simeq\frac{1}{9\kappa_q^2}\,\left[\frac{V_{,\,\phi}\,V_{,\,\phi\phi\phi}}{V^{2q}}
+\frac{(V_{,\,\phi})^2}{V^{2q}}-\frac{5q}{2}\frac{(V_{,\,\phi})^2V_{,\,\phi\phi}}{V^{2q+1}}
+\frac{q(q+2)}{2}\frac{(V_{,\,\phi})^4}{V^{2(q+1)}}\right].
$$

\begin{figure}[th]
\includegraphics[width=4.0in,angle=0,clip=true]{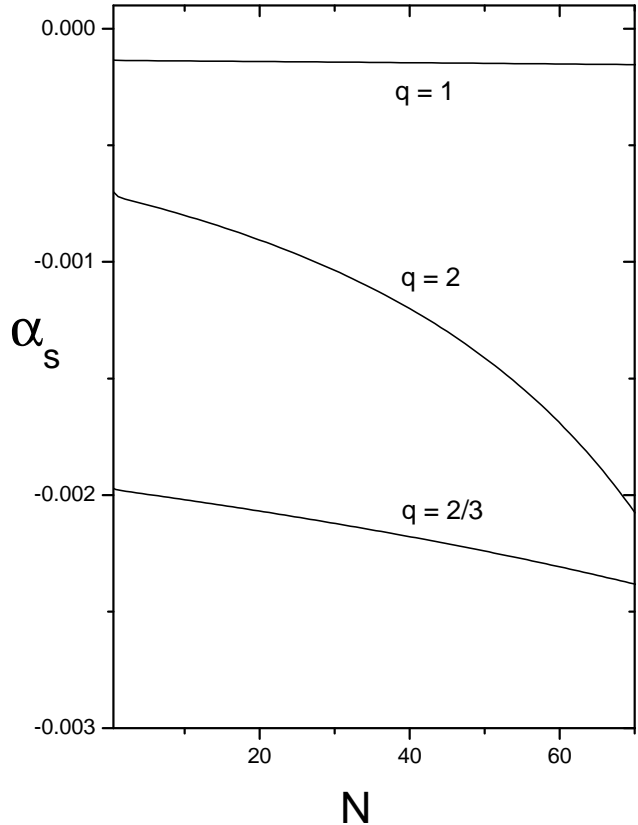}
\caption{This plot shows the running spectral index, $\alpha_s$,
as a function of the N e-folds, for the case $n=2$. We have taken
$\sigma=10^{-10}$, $\kappa_{2/3}=10^{-3}$ and used $m_p=1$.}
\label{fig3}
\end{figure}
\begin{figure}[th]
\includegraphics[width=4.0in,angle=0,clip=true]{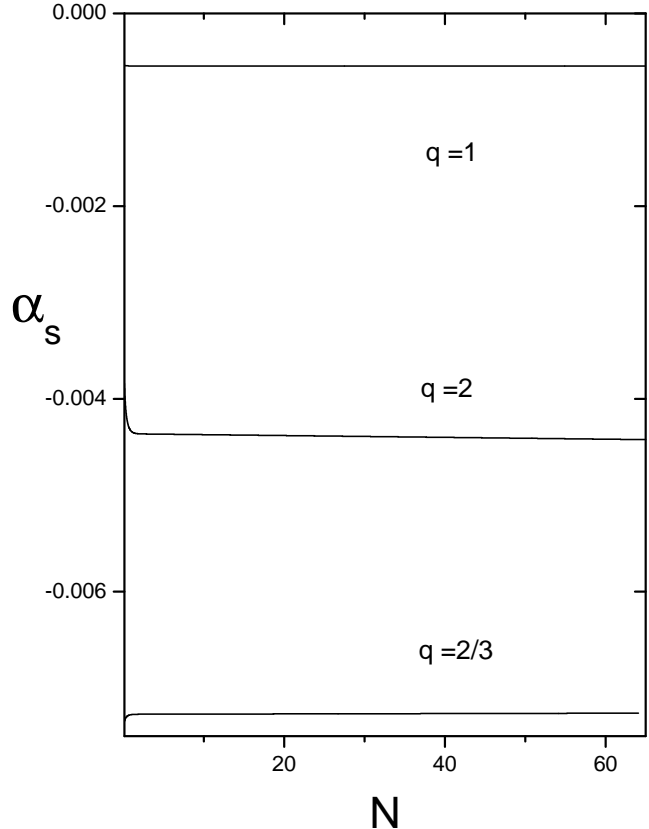}
\caption{Plot shows the running spectral index  as a function of
the N e-folds, for the case $n=4$. We have taken
$\sigma=10^{-10}$, $\kappa_{2/3}=10^{-3}$ and using $m_p=1$.}
\label{fig3}
\end{figure}
\newpage
In models with only scalar fluctuations, the marginalized
derivative value of the spectral index is approximated by
$dn_s/d\ln k=\alpha_s \sim -0.03$ for  Wilkinson Microwave
Anisotropy Probe (WMAP) five-year data only \cite{WMAP3}.

Actually, with this data, together with Sloan Digital Sky Survey
(SDSS) large scale structure surveys \cite{Teg}, an upper bound
for $\alpha_s(k_0$) was found, where $k_0$=0.002 Mpc$^{-1}$
corresponds to $L=\tau_0 k_0\simeq 30$, with the distance to the
decoupling surface $\tau_0$= 14,400 Mpc. SDSS measures galaxy
distributions at red-shifts $a\sim 0.1$ and probes $k$ in the
range 0.016 $h$ Mpc$^{-1}$$<k<$0.011 $h$ Mpc$^{-1}$. The  recent
WMAP five-year data results give the value for  the scalar
curvature spectrum $P_{\cal
R}(k_0)\equiv\,25\delta_H^2(k_0)/4\simeq 2.4\times\,10^{-9}$ .
This value allows  us to find  constraints on the  parameters
appearing in  our model.  In particular, for  $N=10$, we have
$\delta_H \approx 5.6\times 10^{-4}$ for $q=1$, $\delta_H \approx
1.6\times 10^{-4}$ for $q=2$, and $\delta_H \approx 7.2\times
10^{-5}$ for $q=2/3$.  In the particular case when $n=2$, the
running of the scalar spectral index $dn_s/d\ln k=\alpha_s$  are
given by $\alpha_s\sim -1.4\times10^{-4}$ for $q=1$, $\alpha_s\sim
-8.0\times10^{-4}$ for $q=2$, and $\alpha_s\sim -0.002$ for
$q=2/3$. For $n=4$, it is found $\alpha_s\sim -5.4\times10^{-4}$
for $q=1$, $\alpha_s\sim -0.004$ for $q=2$, and $\alpha_s\sim
-0.007$ for $q=2/3$.

\section{Conclusion and Final Remarks}

In this work we have studied  a closed inflationary  universe
model in which the gravitational effects are described by the
Gauss-Bonnet Brane World Cosmology. We study this model by using
the scheme of patch cosmology. In this approach the dynamics of
the scale factor is governed by a modified Friedmann equation
given by $H^2=\kappa_q\,\rho^q-1/a^2$, where $q=1$ represent GR
theory, $q=2$ describes high energy limit of brane world
cosmology, and $q=2/3$ corresponds to brane world cosmology with a
Gauss-Bonnet correction in the bulk.  We have described different
cosmological models where the matter content is given by a single
scalar field in presence of a constant and a self interacting
chaotic potentials.

In the former scenario, we consider a potential with two regimes,
one in which the potential is constant and another, where it
sharply rises to infinity. In the context of Einstein's theory of
GR ($q=1$), this models was study by Linde\cite {Linde:2003hc},
who showed that this model is not optimistic due to the constancy
of the potential implying that the universe collapses very soon or
inflates forever. In the other cases, where  $q=2$, and $q=2/3$,
we could  fix the graceful exit problem due to that in both cases
we add extra ingredients; that is the model depends  on the value
of the brane tension $\sigma$ and the Gauss-Bonnet coupling
constant $\alpha\sim \kappa_{2/3}^{-1}$. This allows us to reach
the value $\phi=0$, which is needed in order to solve the problem.
 However,  another problem is found related to reheating.
This is due to the fact that the scalar field does not oscillate,
and thus we could not recover the Big-Bang scenario (see
Fig.~\ref{fig1}).

On the other hand, in the chaotic inflationary models, $V (\phi) =
\lambda_n \,\phi^n$, the graceful exit and  the reheating problems
are fixed. Therefore, chaotic  models  represent a scenario
adequate  for describing  inflationary mechanisms together with
their consequences.

We have found that the modifications of the Friedmann equations
change some of the characteristic of the spectrum of scalar and
running scalar index. In particular, for $N=10$, we have $\delta_H
\approx 5.6\times 10^{-4}$ for $q=1$, $\delta_H \approx 1.6\times
10^{-4}$ for $q=2$, and $\delta_H \approx 7.2\times 10^{-5}$ for
$q=2/3$.  In the particular case when $n=2$, the running in the
scalar spectral index $dn_s/d\ln k=\alpha_s$  is given by
$\alpha_s\sim -1.4\times10^{-4}$ for $q=1$, $\alpha_s\sim
-8.0\times10^{-4}$ for $q=2$, and $\alpha_s\sim -0.002$ for
$q=2/3$. For $n=4$, the running in the scalar spectral index
$\alpha_s\sim -5.4\times10^{-4}$ for $q=1$, $\alpha_s\sim -0.004$
for $q=2$, and $\alpha_s\sim -0.007$ for $q=2/3$. Certainly the
differences obtained in three different models under study in the
characteristic of the spectrum of scalar and running scalar index
can give clues of higher dimensional theory in particular
deviations from standard results, our main results are summarized
in plots 2, 3 and 4. Finally, we would like to point out that our
Eq.(\ref{ec10}) is an approximation and must be supplemented by
several different contributions in the context of a closed
inflationary universe \cite{Linde:2003hc}. However, one may expect
that the flat-space expression gives a correct result for $N>3$.
We will postpone this important matter at the present time.

\begin{acknowledgments}

This work was supported by COMISION NACIONAL DE CIENCIAS Y
TECNOLOGIA through FONDECYT Grant N$^{0}$.  1070306 (SdC), 1090613
(RH) and 11060515 (JS). It was partially supported by PUCV DI-PUCV
2009. P. L is supported by Direcci\'on de Investigaci\'on de la
Universidad del B\'{\i}o-B\'{\i}o through the Grant N$^{0}$ 096207
1/R.
\end{acknowledgments}


\begin{thebibliography}{99}

\bibitem{IC} A. Guth, Phys. Rev. D {\bf 23}, 347 (1981);
A. Albrecht and P. J. Steinhardt, Phys. Rev. Lett. {\bf 48}, 1220
(1982). ; A. D. Linde, Phys. Lett. {\bf 108B}, 389 (1982), Phys.
Lett. {\bf 129B}, 177 (1983).

\bibitem{libro} A. D. Linde, \emph{Particle Physics and Inflationary
Cosmology}. Harwoord, Chur, Switzerland, (1990).
arXiv:hep-th/0503203.


\bibitem{Peiris:2003ff} H.~V.~Peiris {\it et al.},
Astrophys.\ J.\ Suppl.\  {\bf 148}, 213 (2003);   D.~N.~Spergel
{\it et al.} [WMAP Collaboration], Astrophys.\ J.\ Suppl.\  {\bf
148}, 175 (2003); L. Page  et al. Astrophys.\ J.\ Suppl.\  {\bf
170}, 335 (2007); D.~N.~Spergel {\it et al.}, Astrophys.\ J.\
Suppl.\  {\bf 170}, 377 (2007).

\bibitem{WMAP3} J.~Dunkley {\it et al.}  [WMAP Collaboration],
  %``Five-Year Wilkinson Microwave Anisotropy Probe (WMAP) Observations:
  %Likelihoods and Parameters from the WMAP data,''
  Astrophys.\ J.\ Suppl.\  {\bf 180}, 306 (2009);  G.~Hinshaw {\it et al.},
  %``Five-Year Wilkinson Microwave Anisotropy Probe (WMAP) Observations: Data
  %Processing, Sky Maps, and Basic Results,''
  Astrophys.\ J.\ Suppl.\  {\bf 180}, 225 (2009).
%%%%%%%%%%%%%%%%%%%%% open or closed models %%%%%%%%%%%%%%%%%


\bibitem{op1} A. Linde, Phys. Rev.  D {\bf59}, 023503 (1999).

\bibitem{op} S. del Campo, R. Herrera;
Phys. Rev. D {\bf67}, 063507 (2003);  S. del Campo, R. Herrera, J.
Saavedra, Phys. Rev. D \textbf{70}, 023507 (2004);   L. Balart, S.
del Campo, R. Herrera, P. Labra\~{n}a, J. Saavedra, Phys. Lett. B
\textbf{647}, 313 (2007);  S.~del Campo, R.~Herrera and
J.~Saavedra, Mod.\ Phys.\ Lett.\  A {\bf 23}, 1187 (2008).


\bibitem{Linde:2003hc} A.~Linde, JCAP {\bf 0305}, 002 (2003).

\bibitem{Joel} S. del Campo, R. Herrera, J. Saavedra, Int.\ J.\ Mod.\ Phys.\  D {\bf
14}, 861 (2005).

\bibitem{Cl} S. del Campo and R. Herrera, Class. Quant. Grav.
\textbf{ 22}, 2687 (2005); L. Balart, S. del Campo, R. Herrera, P.
Labra\~{n}a, Eur.\ Phys.\ J.\ C {\bf 51}, 185 (2007).

%%%%%%%%%%%%%%%%%%%%%%%%%%%%%%%%%%%%%%%%%%%%%%%%%%%%%%%%%%%%%%%%%%%%

\bibitem{1}J.E.Lidsey, astro-ph/0305528; J.E.Lidsey, D.Wands and
E.J.Copeland, Phys. Rep. \textbf{337}, 343 (2000); M.Gasperini and
G.Veneziano, Phys. Rep. \textbf{373}, 1 (2003); M.Quevedo, Class.
Quant. Grav. \textbf{19}, 5721 (2002). %[hep-th/0210292].

\bibitem{2} N.Arkani-Hamed, S.Dimopoulos and G.Dvali, Phys. Lett. B
\textbf{429}, 263 (1998); I.Antoniadis, N.Arkani-Hamed,
S.Dimopoulos and G.Dvali, Phys. Lett. B \textbf{436}, 257 (1998);
L.Randall and R.Sundrum, Phys. Rev. Lett. \textbf{83}, 3370
(1999).

%%%%%%%  nuevos 1 %%%%%%%%%%%

\bibitem{witten}  P. Horava and E. Witten, Nucl.Phys.B {\bf 475}, 94 (1996).

\bibitem{witten2}P. Horava and E. Witten, Nucl.Phys.B {\bf 460}, 506 (1996).


\bibitem{lecturer}  J. Lidsey, Lect.\ Notes Phys.\  {\bf 646}, 357
(2004).

\bibitem{lecturer2} P. Brax, C. van de Bruck. Class.Quant.Grav.{\bf 20}, R201-R232
(2003).
\bibitem{lecturer3} E. Papantonopoulos, Lect.Notes Phys. {\bf 592}, 458
(2002).

\bibitem{maartens}  R. Maartens, D. Wands, B. A. Bassett, I. Heard,
Phys.Rev.D {\bf 62}, 041301 (2000); P.~Binetruy, C.~Deffayet,
U.~Ellwanger and D.~Langlois,
 % %``Brane cosmological evolution in a bulk with cosmological constant,''
  Phys.\ Lett.\  B {\bf 477}, 285 (2000);  E.~E.~Flanagan, S.~H.~H.~Tye and I.~Wasserman,
  %``Cosmological expansion in the Randall-Sundrum brane world scenario,''
  Phys.\ Rev.\  D {\bf 62}, 044039 (2000).

%%%%%%%%%%%%%%  nuevos 2%%%%%%%%%%%

\bibitem{Meng:2003pn}  X.~H.~Meng and P.~Wang,
  %``Inflationary attractor in Gauss-Bonnet brane cosmology,''
  Class.\ Quant.\ Grav.\  {\bf 21}, 2527 (2004)
 \bibitem{Lidsey:2003sj}
  J.~E.~Lidsey and N.~J.~Nunes,
  %``Inflation in Gauss-Bonnet brane cosmology,''
  Phys.\ Rev.\  D {\bf 67}, 103510 (2003).



\bibitem{Calcagni:2004as}G.~Calcagni and S.~Tsujikawa,
Phys.\ Rev.\  D {\bf 70}, 103514 (2004).




\bibitem{Kim:2004gs}
  H.~Kim, K.~H.~Kim, H.~W.~Lee and Y.~S.~Myung,
  %``Inflation parameters from Gauss-Bonnet braneworld,''
  Phys.\ Lett.\  B {\bf 608}, 1 (2005).
\bibitem{Kim:2004jc}
  K.~H.~Kim and Y.~S.~Myung,
  %``Cosmological constraints from Gauss-Bonnet braneworld with large-field
  %potentials,''
  JCAP {\bf 0412}, 004 (2004).
\bibitem{Kim:2004hu}
  K.~H.~Kim and Y.~S.~Myung,
  %``Test of patch cosmology with WMAP,''
  Int.\ J.\ Mod.\ Phys.\  D {\bf 14}, 1813 (2005).
\bibitem{Murray:2006fw}
  B.~M.~Murray and Y.~S.~Myung,
  %``Gauss-Bonnet braneworld and WMAP three year results,''
  Phys.\ Lett.\  B {\bf 642}, 426 (2006).



\bibitem{Ch} C.Charmousis and J-F. Dufaux, Class.\ Quant.\ Grav.\  {\bf 19}, 4671
(2002); J.E. Lidsey and N.J. Nunes, Phys. Rev. D, \textbf{67},
103510 (2003); Kei-ichi Maeda, Takashi Torii, Phys.\ Rev.\  D {\bf
69}, 024002 (2004); J.-F. Dufaux, J. Lidsey, R. Maartens, M. Sami,
Phys.\ Rev.\  D {\bf 70}, 083525 (2004);
%
%Shin'ichi Nojiri, Sergei D. Odintsov, hep-th/0006232;
B. Abdesselam and N. Mohammedi, Phys. Rev. D \textbf{65}, 084018
(2002).


\bibitem{Davis:2002gn}
  S.~C.~Davis,
  %``Generalised Israel junction conditions for a Gauss-Bonnet brane world,''
  Phys.\ Rev.\  D {\bf 67}, 024030 (2003).


\bibitem{Gravanis:2002wy}
  E.~Gravanis and S.~Willison,
  %``Israel conditions for the Gauss-Bonnet theory and the Friedmann equation on
  %the brane universe,''
  Phys.\ Lett.\  B {\bf 562}, 118 (2003).


\bibitem{Calcagni:2004bh}
  G.~Calcagni,
  %``Slow-roll parameters in braneworld cosmologies,''
  Phys.\ Rev.\  D {\bf 69}, 103508 (2004).
%  [arXiv:hep-ph/0402126].
  %%CITATION = PHRVA,D69,103508;%%

  \bibitem{Chung:1999zs}
  D.~J.~H.~Chung and K.~Freese,
  %``Cosmological challenges in theories with extra dimensions and remarks  on
  %the horizon problem,''
  Phys.\ Rev.\  D {\bf 61}, 023511 (2000).

\bibitem{Bi}P. Bowcock, C. Charmousis and R. Gregory, Class. Quant. Grav. {\bf17}, 4745 (2000).





\bibitem{Koya} G.~Calcagni,
  %``de Sitter thermodynamics and the braneworld,''
  JHEP {\bf 0509}, 060 (2005).

\bibitem{Fairbairn:2002yp} M.~Fairbairn and M.~H.~G.~Tytgat,
Phys.\ Lett.\ B {\bf 546}, 1 (2002).

\bibitem{Teg} M. Tegmark  et al.,   Phys. Rev. D {\bf 69},
103501 (2004).



\end{thebibliography}
\end{document}